\documentclass[pdflatex,sn-mathphys-num]{sn-jnl}% Math and Physical Sciences Numbered Reference Style

\usepackage{graphicx}%
\usepackage{multirow}%
\usepackage{amsmath,amssymb,amsfonts}%
\usepackage{amsthm}%
\usepackage{mathrsfs}%
\usepackage[title]{appendix}%
\usepackage{xcolor}%
\usepackage{textcomp}%
\usepackage{manyfoot}%
\usepackage{booktabs}%
\usepackage{algorithm}%
\usepackage{algorithmicx}%
\usepackage{algpseudocode}%
\usepackage{listings}%

\theoremstyle{thmstyleone}%

\theoremstyle{thmstyletwo}%

\theoremstyle{thmstylethree}%

\begin{document}

\title[ADD-Net+]{ADD-Net+: A Leakage-Free and Explainable Deep Learning Framework for Alzheimer's Disease Staging from MRI}

\author*[1]{\fnm{Timothy Oluwapelumi} \sur{Adeyemi}}
\email{atimothy.dev@gmail.com}

\affil*[1]{%
\orgname{WeAreGenius Research Institute},
\orgaddress{\city{Lagos} \country{Nigeria}}
}

\abstract{
Alzheimer's disease (AD) staging from MRI is often benchmarked against ADD-Net, whose evaluation protocol applies SMOTETomek resampling before train/validation/test partitioning. This ordering creates a risk of information leakage across splits that may lead to optimistic estimates of model generalisation. We introduce \textbf{ADD-Net+}, a leakage-free, attention-enhanced, explainable convolutional framework that enforces stratified partitioning before any resampling, embeds Squeeze-and-Excitation attention, and integrates Grad-CAM, t-SNE, Expected Calibration Error (ECE), and Brier score as intrinsic evaluation components rather than post-hoc additions. On a four-class MRI benchmark, ADD-Net+ achieves \textbf{89.69\%} accuracy, \textbf{98.75\%} macro-AUC, perfect ModerateDemented recall, an ECE of \textbf{0.0841}, and a Brier score of \textbf{0.1531}. A four-arm ablation shows this gain stems chiefly from the optimisation regime rather than attention alone (55.73\% versus 55.62\%), and that replicating the original resample-before-split protocol reaches only \textbf{64.01\%}, well below the originally reported 98.63\%. Rather than claiming one decisive improvement, ADD-Net+ offers a transparently decomposed, reproducible, and interpretable benchmark for reliable computer-aided AD staging.
}


\maketitle

\section{Introduction}

Alzheimer's disease (AD) is the most common cause of dementia and one of the leading causes of disability among older adults worldwide \citep{Mohsen2025}. Characterised by progressive neurodegeneration, the disease gradually impairs memory, cognitive function, behaviour, and the ability to perform daily activities, imposing substantial emotional, social, and economic burdens on patients, caregivers, and healthcare systems. As global life expectancy continues to increase, the prevalence of Alzheimer's disease is projected to rise significantly, reinforcing the urgent need for reliable diagnostic tools capable of supporting timely clinical intervention. Since therapeutic strategies are generally more effective during the early stages of disease progression, accurate staging remains fundamental to improving patient management and long-term clinical outcomes.

Structural magnetic resonance imaging (MRI) has emerged as one of the most widely adopted neuroimaging modalities for Alzheimer's disease assessment because it enables the non-invasive visualisation of anatomical changes associated with neurodegeneration, including hippocampal atrophy, cortical thinning, and ventricular enlargement \citep{Saltarelli2025}. Nevertheless, the interpretation of MRI scans requires considerable clinical expertise and remains susceptible to inter-observer variability, particularly during the early stages of the disease where structural abnormalities are often subtle. Consequently, automated computer-aided diagnostic systems have attracted increasing research interest as a means of improving diagnostic consistency, reducing clinical workload, and supporting more objective disease assessment \citep{Mienye2025}.

Recent advances in deep learning, particularly convolutional neural networks (CNNs), have substantially improved automated medical image analysis by learning hierarchical feature representations directly from imaging data. Numerous studies have reported promising performance in Alzheimer's disease classification using CNNs, transfer learning, hybrid architectures, and transformer-based models \citep{Abuowaida2025, Hechkel2026}. Despite these advances, several methodological challenges continue to limit their translation into routine clinical practice. Many existing studies employ experimental protocols that inadvertently introduce information leakage during data preprocessing, resulting in overly optimistic performance estimates that do not reflect real-world generalization \citep{Fareed2022}. Furthermore, severe class imbalance frequently biases learning towards majority classes, reducing sensitivity to early-stage Alzheimer's disease, while the limited interpretability of many deep learning models hinders clinical trust and adoption. These challenges highlight the need for diagnostic frameworks that not only achieve high predictive performance but also demonstrate methodological rigor, transparency, and clinical reliability.

To address these limitations, this study proposes \textbf{ADD-Net}, a deep learning framework that enhances the leakage-free attention for multi-class Alzheimer's disease staging using structural magnetic resonance imaging, based on and methodologically correcting the original ADD-Net formulation \citep{Fareed2022}. Unlike many existing approaches, the proposed framework adopts a rigorously designed experimental pipeline that preserves the integrity of model validation by eliminating information leakage throughout the learning process. In addition, attention-guided feature learning is incorporated to improve the representation of disease-specific anatomical characteristics, while explainable artificial intelligence techniques provide visual evidence that supports the interpretability and trustworthiness of model predictions. Collectively, these design choices aim to improve both predictive performance and clinical confidence in automated Alzheimer's disease staging.

The principal contributions of this study are summarized as follows:

\begin{itemize}
    \item A leakage-free experimental framework that preserves the integrity of model validation by preventing information leakage during data preprocessing and model development.
    
    \item An attention-enhanced deep learning architecture, termed \textbf{ADD-Net}, designed to improve discriminative feature learning for robust multi-class Alzheimer's disease staging from structural MRI.
    
    \item An integrated explainability framework combining Grad-CAM, feature-space visualisation, and confidence analysis to improve the transparency and clinical interpretability of model predictions.
    
    \item A comprehensive experimental evaluation demonstrating robust diagnostic performance under realistic imbalanced data conditions, providing a reproducible and trustworthy framework for computer-aided Alzheimer's disease diagnosis.
\end{itemize}

The remainder of this paper is organised as follows. Section~\ref{sec:related} reviews the related literature on deep learning for Alzheimer's disease diagnosis. Section~\ref{sec:methodology} presents the proposed ADD-Net framework and experimental methodology. Section~\ref{sec:results} reports and analyses the experimental results. Section~\ref{sec:discussion} discusses the clinical implications, strengths, and limitations of the proposed framework, while Section~\ref{sec:conclusion} concludes the paper and outlines directions for future research.

\section{Related Work}
\label{sec:related}

The application of artificial intelligence (AI) to Alzheimer's disease (AD) diagnosis has advanced rapidly over the past decade, transitioning from conventional machine learning approaches based on handcrafted feature engineering to end-to-end deep learning (DL) frameworks capable of learning hierarchical representations directly from neuroimaging data. Driven by the increasing availability of large-scale neuroimaging repositories and advances in computational modelling, convolutional neural networks (CNNs), transfer learning strategies, and, more recently, vision transformer architectures have demonstrated substantial improvements in identifying complex structural biomarkers from magnetic resonance imaging (MRI) for automated Alzheimer's disease diagnosis \cite{Mienye2025, Mohsen2025, Raza2025, Jo2019}. Despite these advances, the clinical translation of these models remains constrained by persistent challenges relating to methodological rigour, reliable multi-class disease staging, model interpretability, and vulnerability to information leakage during experimental evaluation \cite{Afreen2025, Saltarelli2025, Kantarci2025}.

\subsection{Deep Learning and Multi-Class Staging in Alzheimer's Disease}

Deep learning has become the dominant paradigm for automated Alzheimer's disease diagnosis owing to its ability to learn discriminative imaging features directly from raw neuroimaging data without relying on handcrafted descriptors. While early studies primarily focused on binary classification tasks, such as distinguishing Alzheimer's disease from cognitively normal subjects, clinical decision-making requires fine-grained multi-class staging to accurately monitor disease progression and facilitate timely intervention. Addressing this challenge, Kazemi et al. \cite{Kazemi2018} demonstrated the feasibility of multi-class disease staging by employing the AlexNet architecture on rigorously preprocessed functional MRI (fMRI) data to classify five cognitive stages, including normal cognition (NC), significant memory concern (SMC), early mild cognitive impairment (EMCI), late mild cognitive impairment (LMCI), and Alzheimer's disease (AD). Similarly, Murugan et al. \cite{Murugan2021} proposed DEMNET, a lightweight CNN architecture capable of accurately distinguishing four stages of dementia from structural MRI, highlighting the potential of computationally efficient deep learning models for practical clinical deployment.

A major challenge associated with multi-class Alzheimer's disease classification is the severe class imbalance commonly observed in clinical neuroimaging datasets, which often biases model learning towards majority disease categories. To mitigate this limitation, recent studies have incorporated advanced resampling strategies, particularly the Synthetic Minority Over-sampling Technique (SMOTE), to improve minority-class representation, reduce overfitting, and enhance diagnostic sensitivity across all disease stages \cite{Murugan2021}. Beyond addressing class imbalance, researchers have increasingly explored multimodal learning strategies that integrate complementary information from structural MRI, electronic health records, and genomic single nucleotide polymorphisms (SNPs). For example, Venugopalan et al. \cite{Venugopalan2021} demonstrated that combining multimodal data through three-dimensional CNNs and stacked autoencoders provides richer feature representations and superior staging performance compared with single-modality approaches. Likewise, although non-neuroimaging modalities such as smartphone accelerometer mobility signals have shown promise for monitoring disease progression using one-dimensional CNNs \cite{Bringas2020}, structural and functional MRI continue to represent the most reliable imaging modalities for capturing clinically relevant neurodegenerative biomarkers.

Despite these significant advances, important limitations remain. Many recently proposed architectures primarily optimise predictive accuracy on benchmark datasets without adequately addressing generalisability across heterogeneous and clinically imbalanced populations \cite{Abuowaida2025, Asaduzzaman2025, Alasiry2025}. Moreover, increasingly sophisticated ensemble and transformer-based models often achieve superior predictive performance at the expense of model transparency, thereby reinforcing the "black-box" nature of deep learning and limiting clinicians' ability to interpret the pathological basis of automated diagnostic decisions \cite{Alruily2025, Khaleel2025, Hechkel2026, Lv2026}.

\subsection{Attention Mechanisms and Explainable Artificial Intelligence}

As deep learning architectures continue to increase in complexity, Explainable Artificial Intelligence (XAI) has evolved from a desirable enhancement to a fundamental requirement for trustworthy clinical decision support. Among the numerous techniques proposed to improve model interpretability, attention mechanisms have proven particularly effective by enabling neural networks to selectively emphasise diagnostically informative anatomical structures, such as the hippocampus and entorhinal cortex, while suppressing irrelevant background information. This adaptive feature refinement not only improves discriminative learning but also enhances the clinical relevance of extracted representations \cite{Fareed2022, Abuowaida2025}.

Complementing attention mechanisms, a variety of post-hoc explainability techniques have been introduced to provide visual evidence supporting automated predictions. Murugan et al. \cite{Murugan2021} employed occlusion sensitivity mapping to identify the brain regions contributing most strongly to dementia classification, while Venugopalan et al. \cite{Venugopalan2021} proposed a perturbation-based feature extraction and clustering framework capable of identifying clinically meaningful biomarkers from multimodal deep learning models. Their findings demonstrated strong agreement between learned imaging features and established Alzheimer's disease biomarkers, including hippocampal atrophy and Rey Auditory Verbal Learning Test scores.

Despite these encouraging developments, explainability remains largely confined to post-hoc visualisation rather than being embedded within the model development process itself. The original ADD-Net~\cite{Fareed2022} is illustrative of this pattern: its use of Grad-CAM is presented as a qualitative visualisation exercise, without confidence, calibration, or uncertainty analysis, and without any discussion of the reliability or limitations of the saliency evidence produced. Consequently, many existing frameworks provide limited insight into the robustness, reliability, and clinical plausibility of their internal decision-making processes, thereby restricting clinician confidence and impeding the adoption of deep learning systems in routine diagnostic practice \cite{Jo2019, Raza2025}.

\subsection{Methodological Rigour, Data Leakage, and Research Gap}

Although recent deep learning models have reported remarkably high diagnostic performance, methodological rigour remains undermined by an often overlooked challenge: information leakage. As comprehensively discussed by Jo et al. \cite{Jo2019}, inappropriate preprocessing strategies and subject allocation procedures can inadvertently introduce patient-specific information into both the training and testing datasets. Such leakage leads to overly optimistic performance estimates that fail to reflect genuine model generalisation and substantially weaken the reproducibility and clinical credibility of reported findings \cite{Mohsen2025, Raza2025}.

The original ADD-Net~\cite{Fareed2022} provides a directly relevant illustration of this risk on the very benchmark used in this paper. Its methodology describes applying the SMOTETomek algorithm to up-sample the full 6,400-image dataset to 12,800 images \emph{before} the resulting balanced dataset is divided into training, validation, and test partitions. Because SMOTE-generated samples are synthetic interpolations of existing minority-class images and their nearest neighbours, this ordering creates the possibility that synthetic derivatives of a given source image land in the test set while the source image itself (or a different derivative of it) resides in the training set. The model may therefore not be evaluated on fully independent data, which could help explain the compressed gap between training and validation accuracy reported in that work and the exceptionally high aggregate metrics (98.63\% accuracy, 99.76\% AUC) relative to the more modest performance the same architecture achieves on the unbalanced dataset (66.1\% accuracy) before SMOTETomek is applied. We note that we have not reproduced the original authors' exact training run, and some of their hyperparameters are not fully specified in the published paper (Section~\ref{subsec:ablation}), so this account should be read as identifying a methodological risk in the published protocol rather than as a claim about what happened in their specific experiment. A comparable resample-before-split protocol underlies subsequent SMOTE/Tomek-augmented CNNs evaluated on the same benchmark~\cite{Ganesan2026}, suggesting the issue is not isolated to a single study but a recurring pattern in this line of work. Beyond this leakage concern, the original ADD-Net's evaluation is also narrow in scope: it reports accuracy, precision, recall, F1-score, AUC, and loss, but omits balanced-accuracy measures, calibration analysis, confidence estimation, and any quantitative treatment of explainability, and its stated limitations are confined to a brief acknowledgement that performance is imbalance-sensitive.

Collectively, the existing literature, and the original ADD-Net in particular, reveals four important limitations that continue to hinder the development of clinically reliable Alzheimer's disease staging systems:

\begin{enumerate}
    \item \textbf{Information Leakage:} Many existing preprocessing and evaluation protocols, including that of the original ADD-Net~\cite{Fareed2022}, resample before partitioning, which risks incompletely isolating image-level information between splits and can compromise the integrity of model validation and lead to optimistic estimates of reported performance.
    
    \item \textbf{Class Imbalance:} Multi-class diagnostic frameworks frequently exhibit biased predictions towards majority disease categories because class imbalance is not addressed in a sufficiently robust and principled manner.
    
    \item \textbf{Limited Explainability:} Existing explainability techniques, including the Grad-CAM analysis in the original ADD-Net, are predominantly applied as post-hoc visualisation tools rather than being intrinsically integrated into the feature learning and decision-making processes, with no accompanying calibration or confidence analysis.
    
    \item \textbf{Fragmented Framework Design:} Current studies rarely combine leakage-free evaluation, attention-guided feature learning, and explainable artificial intelligence within a unified framework for robust multi-class Alzheimer's disease staging.
\end{enumerate}

Motivated by these limitations, and in direct response to the specific leakage risk identified in ADD-Net~\cite{Fareed2022}, this study proposes \textbf{ADD-Net+}, a leakage-free attention-enhanced explainable deep learning framework for multi-class Alzheimer's disease staging using structural MRI. Building on the original ADD-Net's convolutional design philosophy, the proposed framework preserves the integrity of model validation through strict image-level data partitioning performed \emph{before} any resampling, integrates attention mechanisms directly into the feature learning pipeline, and incorporates explainable artificial intelligence as an inherent component of model development rather than a post-hoc addition. Collectively, these design choices establish a robust, transparent framework that provides a more reliable benchmark for computer-aided Alzheimer's disease staging, and a more defensible performance baseline for this benchmark going forward.

\section{Materials and Methods}
\label{sec:methodology}

This section presents the design and implementation of \textbf{ADD-Net+}, a leakage-free, attention-enhanced, and explainable deep learning framework for multi-class Alzheimer's disease staging from structural MRI. The framework integrates dataset preparation, leakage-free preprocessing and class-imbalance correction, an attention-enhanced architecture, a convergence-aware training strategy, and an embedded explainability suite within a single reproducible pipeline. Table~\ref{tab:improvements} (Section~\ref{subsec:architecture}) summarises, point by point, how each design decision in ADD-Net+ directly corrects a specific methodological weakness identified in the original ADD-Net~\cite{Fareed2022} in Section~\ref{sec:related}.

\subsection{System Overview and Design Philosophy}
\label{subsec:overview}

ADD-Net+ is organised around two guiding principles: diagnostic accuracy and transparent interpretability, pursued jointly rather than as separate concerns. It retains the compact, from-scratch convolutional design philosophy of the original ADD-Net~\cite{Fareed2022}, a network with relatively few parameters, well suited to a modestly sized MRI dataset, while substantively revising the experimental pipeline around it: Squeeze-and-Excitation (SE) attention is embedded within the convolutional backbone, class imbalance is corrected without leaking information across data partitions, and an embedded explainability suite is built into the pipeline from the outset.

\begin{figure}[t]
    \centering
    \includegraphics[width=0.85\textwidth]{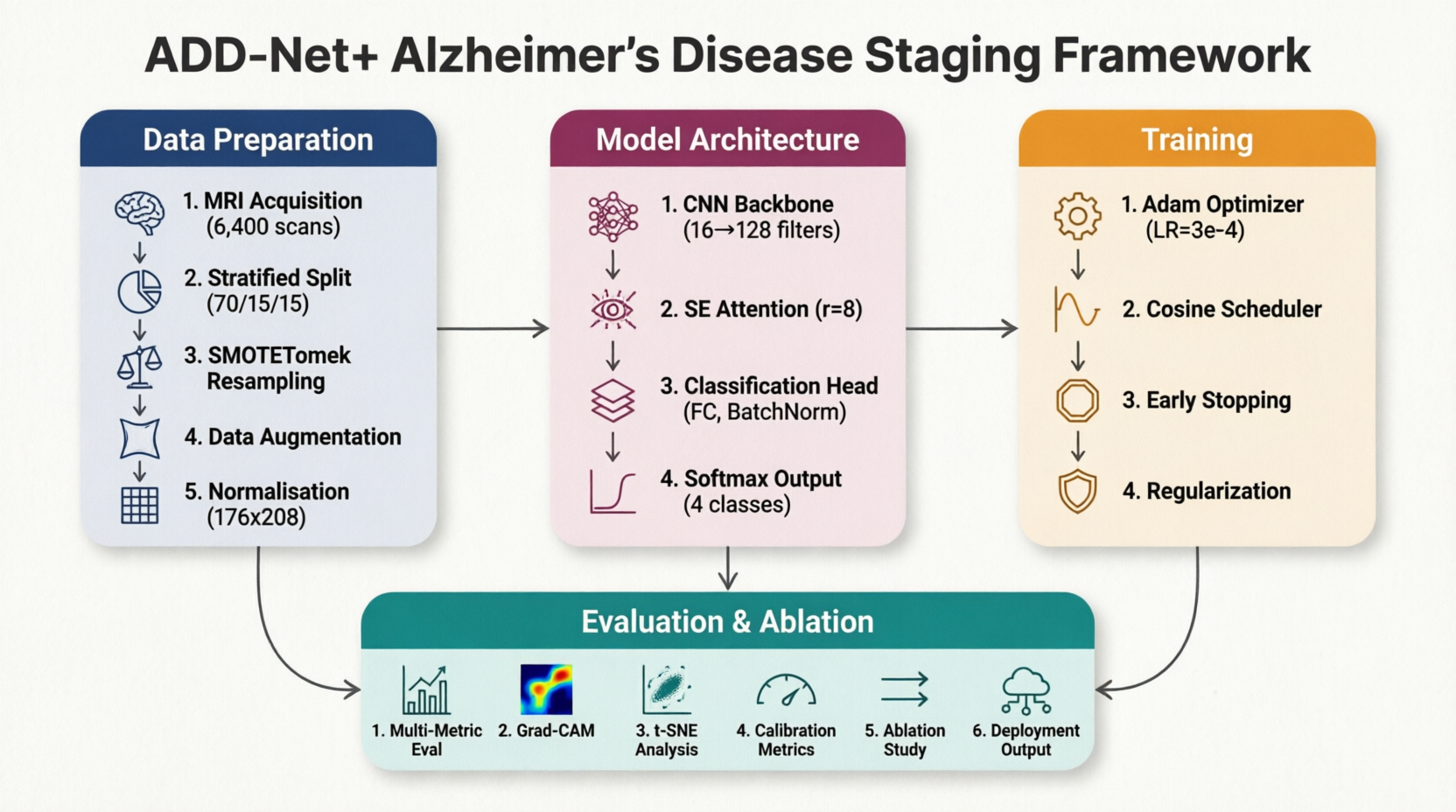}
    \caption{End-to-end ADD-Net+ pipeline, from raw MRI acquisition to explainable clinical output.}
    \label{fig:architecture}
\end{figure}

As illustrated in Figure~\ref{fig:architecture}, the pipeline follows three sequential phases. \emph{Phase~I} verifies, normalises, and stratifies the MRI scans into training, validation, and test partitions \emph{first}, with SMOTETomek resampling subsequently applied strictly to the training partition, the reverse ordering of the original ADD-Net's protocol, and the change that most directly prevents information leakage into held-out data. \emph{Phase~II} trains a custom convolutional backbone augmented with SE attention modules using the Adam optimiser with cosine-annealed warm restarts and label smoothing, the latter improving probability calibration on synthetically generated samples. \emph{Phase~III} quantifies model performance using standard classification metrics while Grad-CAM saliency maps and t-SNE feature-space embeddings provide complementary spatial and representational insight into model behaviour.

\subsection{Dataset and Clinical Context}
\label{subsec:dataset}

The empirical basis of this study is the same publicly available Alzheimer's Disease Neuroimaging MRI dataset used to evaluate the original ADD-Net~\cite{Fareed2022}, comprising 6{,}400 axial T1-weighted MRI slices at a native resolution of $176\times208$ pixels. Using the same dataset as the baseline being corrected allows the effect of the leakage-free protocol to be assessed on a like-for-like basis, isolating methodological differences from dataset differences. Prior to release, all scans underwent standard neuroimaging preprocessing, including skull-stripping, intensity normalisation, and spatial registration to a common anatomical template. Each scan is labelled according to the Clinical Dementia Rating (CDR) scale. Table~\ref{tab:dataset} summarises the class distribution and clinical correlates of each stage. The publicly released dataset does not provide patient or subject identifiers, so partitioning in this study, like in the original ADD-Net, is necessarily performed at the image level rather than the patient level; consequently, while the leakage-free protocol described below eliminates leakage of \emph{synthetic} samples across splits, the possibility that multiple slices from the same individual appear in both training and evaluation partitions cannot be fully excluded from the source data alone, and this is noted as a limitation in Section~\ref{subsec:limitations}.

\begin{table}[!ht]
    \centering
    \caption{Dataset composition and CDR-aligned clinical definitions.}
    \label{tab:dataset}
    \begin{tabular}{@{}lrrrp{5.2cm}@{}}
        \toprule
        \textbf{Dementia stage} & \textbf{$n$} & \textbf{Proportion} & \textbf{CDR} & \textbf{Clinical description} \\
        \midrule
        NonDemented       & 3{,}200 & 50.0\% & 0.0 & Cognitively normal; no detectable impairment \\
        VeryMildDemented  & 2{,}240 & 35.0\% & 0.5 & Subtle memory deficits; daily functioning preserved \\
        MildDemented      & 896     & 14.0\% & 1.0 & Mild impairment progressing toward early-stage AD \\
        ModerateDemented  & 64      & 1.0\%  & 2.0 & Significant cognitive decline; assistance required for daily activities \\
        \bottomrule
    \end{tabular}
\end{table}

A defining characteristic of the dataset is its severe class imbalance. The imbalance ratio between the majority NonDemented class and the minority ModerateDemented class is
\begin{equation}
    \mathrm{IR} = \frac{N_{\text{majority}}}{N_{\text{minority}}} = \frac{3200}{64} = 50,
    \label{eq:ir}
\end{equation}
which substantially exceeds the IR $>10$ threshold conventionally regarded as severe~\cite{Fareed2022}. This distribution reflects genuine clinical prevalence rather than a sampling artefact, consistent with real-world screening populations in which early or asymptomatic presentations vastly outnumber advanced disease. A clinically useful staging model must therefore retain high sensitivity to subtle, early-stage abnormalities, rather than optimising solely for aggregate accuracy; this motivates the resampling and training strategies adopted in Sections~\ref{subsec:preprocessing} and~\ref{subsec:training}. We note, however, that preserving clinical realism introduces an unavoidable trade-off between statistical representativeness and minority-class sensitivity, which cannot be fully resolved by preprocessing alone and is revisited in Section~\ref{sec:discussion}.

\subsection{Data Preprocessing}
\label{subsec:preprocessing}

Raw MRI scans were transformed into a representation suitable for deep learning through a structured pipeline comprising intensity normalisation, stratified partitioning, class-imbalance correction, data augmentation, and label encoding. Unlike the original ADD-Net protocol, in which balancing precedes splitting, every step below is ordered so that partitioning happens first and all subsequent stochastic corrections act on the training partition alone.

\textbf{Image normalisation.} Each scan was loaded at its native resolution ($176\times208\times3$) and converted to a floating-point tensor, with pixel intensities linearly rescaled from $[0,255]$ to $[0,1]$:
\begin{equation}
    I_{\text{norm}}(x,y) = \frac{I_{\text{raw}}(x,y)}{255}.
    \label{eq:norm}
\end{equation}
This stabilises gradient propagation and preserves the relative intensity variation that encodes diagnostically relevant features such as grey-matter density and ventricular enlargement.

\textbf{Stratified partitioning (performed before resampling).} The \emph{original}, un-resampled dataset was divided into training (70\%), validation (15\%), and test (15\%) subsets using stratified sampling with a fixed random seed (42), yielding $N_{\text{train}}=4{,}480$, $N_{\text{val}}=960$, and $N_{\text{test}}=960$ samples. For the minority ModerateDemented class ($n=64$), this corresponds to approximately 45, 10, and 10 samples per split. Because partitioning is performed \emph{before} any synthetic sample is generated, no sample in the validation or test set can share provenance with a synthetic training sample, the specific property absent from the original ADD-Net's protocol. Class imbalance was preserved in the validation and test sets so that reported performance reflects the true clinical distribution.

\textbf{Class-imbalance correction (training partition only).} Preliminary experiments with class-weighted loss proved unstable, with early amplification of minority-class gradients driving representation collapse and near-zero F1 scores on rare classes. Hybrid SMOTETomek resampling was therefore adopted, combining SMOTE oversampling with Tomek-link removal, the same algorithm used in the original ADD-Net, but here applied \emph{only after} partitioning and \emph{only to} the training subset. Synthetic minority samples are generated by interpolation between a sample and one of its $k=3$ nearest neighbours,
\begin{equation}
    x_{\text{new}} = x_i + \lambda (x_{\text{nn}} - x_i), \qquad \lambda \sim U(0,1),
    \label{eq:smote}
\end{equation}
while Tomek-link removal discards ambiguous majority-class samples lying near decision boundaries, improving separability between clinically adjacent stages (e.g.\ VeryMildDemented versus MildDemented). % TODO: confirm the exact post-SMOTETomek training-set size and per-class counts against the printed run log -- Tomek-link removal makes the post-resampling distribution only approximately, not exactly, balanced.
Resampling increased the training set from 4{,}480 images to a near-balanced set (approximately 2{,}240 per class); images were temporarily flattened for resampling and reshaped thereafter, with all stochastic operations fixed under seed 42. Because Tomek-link removal leaves a small residual imbalance, class weights (computed via scikit-learn's balanced heuristic and clipped to $[0.5, 3.0]$) were additionally applied during training to prevent this residual imbalance from dominating the gradient signal. Critically, and in direct contrast to the original ADD-Net protocol, resampling was applied exclusively to the training set; validation and test distributions were left unaltered, preserving both clinical realism and the leakage-free evaluation protocol emphasised throughout this study.

\textbf{Data augmentation.} A medically informed stochastic augmentation scheme was applied to the training set at runtime via a parallelised data pipeline: random horizontal flipping ($p=0.5$) exploits approximate bilateral brain symmetry; brightness ($\pm15\%$) and contrast ($[0.85,1.15]$) jitter mimic scanner-specific gain and pulse-sequence variability; and random cropping with 20-pixel padding (restored to $176\times208$) improves robustness to minor field-of-view differences across scanners. Augmentation was restricted to the training data; validation and test images were left unmodified.

\textbf{Label encoding.} Class labels were one-hot encoded, with each of the four dementia stages represented as a distinct binary vector, enabling multi-class classification via a softmax output layer.

\subsection{Proposed ADD-Net+ Architecture}
\label{subsec:architecture}

ADD-Net+ is a purpose-built convolutional neural network for multi-class dementia staging from structural MRI, trained from scratch rather than fine-tuned from natural-image weights, allowing it to learn anatomically grounded representations unburdened by extraneous visual priors, the same design rationale as the original ADD-Net~\cite{Fareed2022}. The network maps a normalised $176\times208\times3$ input through four SE-attention-enhanced convolutional blocks, flattening, and two regularised dense layers to a four-class softmax output. Table~\ref{tab:improvements} summarises how each component of ADD-Net+ directly addresses a specific methodological limitation of the original ADD-Net architecture and evaluation protocol.

\begin{table}[t]
    \centering
    \caption{ADD-Net+ improvements over the original ADD-Net~\cite{Fareed2022}, mapped to the specific methodological limitation each one corrects.}
    \label{tab:improvements}
    \begin{tabular}{@{}p{2.6cm}p{4.6cm}p{4.6cm}@{}}
        \toprule
        \textbf{Improvement} & \textbf{Issue in original ADD-Net} & \textbf{ADD-Net+ solution} \\
        \midrule
        Leakage-free resampling & SMOTETomek applied to the full dataset before it was split into train/validation/test, allowing synthetic derivatives of a sample to appear on both sides of the split & Stratified partitioning performed first; resampling restricted strictly to the training set thereafter \\
        Stratified partitioning & 60/20/20 split applied \emph{after} balancing, so reported splits describe the balanced (partly synthetic) dataset rather than the original image counts & 70/15/15 stratified split of the \emph{original} dataset with a fixed seed (42), before any synthetic sample exists \\
        SE attention modules & Plain convolutional blocks with no mechanism for channel-wise feature selectivity & SE modules amplify clinically salient channels and suppress noise \\
        Convergence-aware optimiser & SGD with fixed learning rate; early stopping criterion not tied to a class-imbalance-robust metric & Adam with cosine annealing and warm restarts; early stopping on macro-AUC \\
        Integrated explainability & Grad-CAM used only as a qualitative post-hoc visualisation, with no calibration, confidence, or uncertainty analysis & Grad-CAM, t-SNE, and confidence/calibration analysis embedded in the pipeline \\
        \bottomrule
    \end{tabular}
\end{table}

\textbf{Convolutional backbone.} The backbone comprises four hierarchical convolutional blocks with increasing filter depth (16, 32, 64, 128): shallow layers capture low-level cues such as intensity gradients and tissue boundaries, while deeper layers encode higher-order anatomical patterns, cortical thinning, hippocampal atrophy, and ventricular enlargement. Each block applies convolution, batch normalisation, ReLU activation, SE attention, and average pooling; average pooling is adopted to preserve diffuse anatomical information characteristic of early neurodegeneration. Kernel size decreases progressively across blocks ($7\times7$, $5\times5$, $5\times5$, $3\times3$), supporting a transition from coarse to fine-grained feature extraction.

\textbf{Squeeze-and-Excitation attention.} Each convolutional block is augmented with an SE module for adaptive, channel-wise feature recalibration, the principal architectural addition relative to the original ADD-Net, whose convolutional blocks contained no attention mechanism. A \emph{squeeze} step applies global average pooling to obtain a channel-wise descriptor, followed by an \emph{excitation} step in which a bottleneck network (reduction ratio $r=8$) produces channel-wise weights that rescale the original feature maps:
\begin{equation}
    e = \sigma\!\left(W_2 \cdot \mathrm{ReLU}(W_1 \cdot s + b_1) + b_2\right), \qquad
    \tilde{u}_c = e_c \cdot u_c,
    \label{eq:se}
\end{equation}
where $s$ denotes the squeezed descriptor and $e_c$ modulates channel $c$, selectively amplifying clinically relevant signals while attenuating noise with negligible parameter overhead.

\textbf{Classification head.} Following the final SE-attention block, a dropout layer ($p=0.35$) is applied to the pooled feature maps before flattening into an 18{,}304-dimensional vector. This vector is passed to a two-stage fully connected classifier (256 and 128 units), each followed by batch normalisation and dropout (0.45 and 0.225, respectively) to control overfitting in this high-dimensional space. The network comprises 4{,}868{,}084 trainable parameters in total. A softmax output layer produces a calibrated probability distribution over the four dementia stages, supporting both classification and the confidence-aware review workflow described in Section~\ref{subsec:explainability}.

\subsection{Training and Optimisation Strategy}
\label{subsec:training}

The optimisation strategy addresses three coupled challenges: learning from a severely imbalanced dataset, controlling overfitting induced by synthetic SMOTETomek samples, and ensuring stable yet sufficiently exploratory convergence. ADD-Net+ was trained for a maximum of 80 epochs, subject to the early-stopping criterion described below. Table~\ref{tab:hyperparams} summarises the complete configuration.

\begin{table}[t]
    \centering
    \caption{Optimisation hyperparameter configuration and justification.}
    \label{tab:hyperparams}
    \begin{tabular}{@{}p{3.2cm}p{2cm}p{6.5cm}@{}}
        \toprule
        \textbf{Hyperparameter} & \textbf{Value} & \textbf{Justification} \\
        \midrule
        Initial learning rate & $3\times10^{-4}$ & Balances convergence speed and gradient stability under imbalanced class conditions \\
        Weight decay ($L_2$) & $5\times10^{-5}$ & Moderate regularisation without over-constraining representational capacity \\
        Label smoothing ($\alpha$) & 0.08 & Reduces overconfidence on synthetic SMOTETomek samples; improves probability calibration \\
        Cosine annealing period ($T_0$) & 30 epochs & Provides sufficient exploration within each cycle prior to warm restart \\
        Batch size & 32 & Balances beneficial gradient noise for generalisation against computational efficiency \\
        Early-stopping patience & 20 epochs & Accommodates learning-rate cycling; avoids premature termination during warm-restart plateaus \\
        Dropout (dense 1 / dense 2) & 0.45 / 0.225 & Higher rate where the feature space is largest; reduced in the second layer to preserve discriminative capacity \\
        \bottomrule
    \end{tabular}
\end{table}

\textbf{Loss function.} Training minimises categorical cross-entropy with label smoothing ($\alpha=0.08$), which redistributes a small fraction of probability mass from the true class to all classes:
\begin{equation}
    y_k^{\text{smooth}} = (1-\alpha)\,y_k + \frac{\alpha}{K}, \qquad
    \mathcal{L} = -\sum_{k} y_k^{\text{smooth}} \log(\hat{y}_k),
    \label{eq:labelsmooth}
\end{equation}
with $K=4$ classes, improving probability calibration in clinically ambiguous cases.

\textbf{Optimiser and regularisation.} Model parameters were optimised using Adam ($\eta_0=3\times10^{-4}$, $\beta_1=0.9$, $\beta_2=0.999$) with $L_2$ weight decay ($\lambda=5\times10^{-5}$) to improve convergence stability under class imbalance, in place of the fixed-rate SGD optimiser used in the original ADD-Net.

\textbf{Learning-rate schedule.} A cosine annealing schedule with warm restarts governs the learning rate:
\begin{equation}
    \eta_t = \eta_{\min} + \frac{1}{2}(\eta_{\max} - \eta_{\min})\left(1 + \cos\left(\frac{\pi T_{\text{cur}}}{T_0}\right)\right),
    \label{eq:cosine}
\end{equation}
with $\eta_{\max}=3\times10^{-4}$, $\eta_{\min}=1\times10^{-5}$, and $T_0=30$ epochs, decaying smoothly within each cycle before resetting to promote exploration of new regions of the loss landscape.

\textbf{Early stopping.} Training was regulated by early stopping with a patience of 20 epochs, monitored via macro-averaged ROC-AUC, which is insensitive to class imbalance and provides a more reliable indicator of generalisation than accuracy or loss. The checkpoint with the highest validation AUC was retained for final evaluation.

\subsection{Explainability and Clinical Interpretability}
\label{subsec:explainability}

Predictive performance alone is insufficient for clinical deployment; diagnostic decisions must also be transparent and interpretable. Where the original ADD-Net relied on Grad-CAM alone as a qualitative illustration, ADD-Net+ incorporates three complementary and quantitatively grounded explainability mechanisms.

\textbf{Grad-CAM saliency visualisation.} Gradient-weighted Class Activation Mapping (Grad-CAM) highlights the MRI regions with the greatest influence on a given prediction:
\begin{equation}
    L_{\text{Grad-CAM}}^{c} = \mathrm{ReLU}\!\left(\sum_{k} \alpha_k^{c} A^{k}\right),
    \qquad
    \alpha_k^{c} = \frac{1}{Z}\sum_{i}\sum_{j} \frac{\partial \hat{y}^{c}}{\partial A_{ij}^{k}},
    \label{eq:gradcam}
\end{equation}
where $\hat{y}^{c}$ is the predicted score for class $c$, $A_{ij}^{k}$ denotes the activation of channel $k$ at spatial location $(i,j)$, and $Z$ normalises over spatial dimensions. The resulting heatmap is upsampled to the original resolution and overlaid on the source scan; clinically relevant activations are expected to coincide with recognised biomarkers such as the hippocampus and periventricular regions. Grad-CAM reflects statistical association rather than causal explanation, and is best understood as a decision-support tool for assessing the anatomical plausibility of a prediction.

\textbf{t-SNE feature-space visualisation.} t-distributed stochastic neighbour embedding is applied to the 128-dimensional feature vectors from the penultimate layer, projecting them to two dimensions while preserving local neighbourhood relationships. Distinct, well-separated clusters indicate strong class separability, whereas overlapping clusters reveal ambiguity not captured by aggregate metrics; as t-SNE preserves local but not global structure, inter-cluster distances are treated as qualitative rather than as precise measures of class separation.

\textbf{Confidence analysis.} Softmax outputs are further used to assess predictive uncertainty via the maximum class probability (confidence). The distribution of confidence is compared between correctly and incorrectly classified test cases, and across the four dementia stages; substantial overlap between the confidence distributions of correct and incorrect predictions would indicate elevated misclassification risk, while a clear separation supports the reliability of high-confidence predictions. Low-confidence predictions are natural candidates for clinical review, a consideration of particular importance for adjacent stages such as VeryMildDemented and MildDemented, where inter-class differences are subtle but clinically consequential.

\textbf{Calibration metrics.} Confidence dispersion alone does not establish calibration, a model can be dispersed yet systematically over- or under-confident. We therefore report three standard, quantitative calibration measures in addition to the confidence distributions above:

\begin{itemize}
    \item \emph{Expected Calibration Error (ECE)}~\cite{Guo2017,Naeini2015}: predictions are partitioned into $M=15$ equal-width bins by top-label confidence $\hat{p} = \max_k \hat{y}_k$; within each bin $B_m$, the gap between mean confidence and empirical accuracy is computed and combined as a sample-weighted average,
    \begin{equation}
        \mathrm{ECE} = \sum_{m=1}^{M} \frac{|B_m|}{N} \left| \mathrm{acc}(B_m) - \mathrm{conf}(B_m) \right|,
        \label{eq:ece}
    \end{equation}
    where $N$ is the number of test samples. $\mathrm{ECE} \in [0,1]$, with 0 indicating perfect calibration.

    \item \emph{Multi-class Brier score}~\cite{Brier1950}: the mean squared error between the predicted probability vector $\hat{y}_i \in \mathbb{R}^K$ and the one-hot ground truth $o_i$ across all $N$ test samples and $K=4$ classes,
    \begin{equation}
        \mathrm{BS} = \frac{1}{N} \sum_{i=1}^{N} \sum_{k=1}^{K} \left( \hat{y}_{ik} - o_{ik} \right)^2,
        \label{eq:brier}
    \end{equation}
    with lower values indicating better-calibrated, sharper probability estimates.

    \item \emph{Reliability diagram}: the standard visual companion to ECE, plotting empirical accuracy $\mathrm{acc}(B_m)$ against mean confidence $\mathrm{conf}(B_m)$ for each bin against the diagonal of perfect calibration, giving a qualitative read of whether the model is over-confident, under-confident, or well calibrated across the confidence range.
\end{itemize}

These three measures, together with the confidence-distribution analysis above, are reported in Section~\ref{subsec:calibration} and directly justify the use of the term \emph{calibration} rather than confidence analysis alone.

\subsection{Implementation Environment}
\label{subsec:environment}

ADD-Net+ was implemented in Python using TensorFlow/Keras and trained on a cloud-based Google Colaboratory platform with GPU acceleration (NVIDIA T4 and A100 instances). NumPy and Pandas supported numerical computation and dataset management; Scikit-learn provided stratified partitioning, evaluation metrics, and t-SNE; Imbalanced-learn implemented SMOTETomek resampling; SciPy supported the numerical operations underlying Grad-CAM; and Plotly was used for interactive visualisation of training dynamics and explainability outputs. All stochastic components, dataset splitting, weight initialisation, data augmentation, and SMOTETomek resampling, were controlled with a fixed random seed to ensure reproducibility.

Several methodological trade-offs are acknowledged explicitly for transparency: the potential for residual bias introduced by synthetic oversampling, the qualitative rather than metric nature of t-SNE embeddings, and the non-causal interpretation of Grad-CAM saliency. The four-arm ablation (Section~\ref{subsec:ablation}) and the calibration metrics (Section~\ref{subsec:calibration}) are implemented as parametrised extensions of this same pipeline, reusing the identical data-loading, augmentation, and architecture-construction functions, so that the only factors that vary across ablation arms are exactly those stated in Section~\ref{subsec:ablation}. The following section evaluates ADD-Net+ under leakage-free experimental conditions and compares its diagnostic performance, robustness, calibration, and interpretability against the original ADD-Net and other existing state-of-the-art approaches.

\section{Results}
\label{sec:results}

ADD-Net+ was evaluated on the held-out test partition (960 scans, Section~\ref{subsec:preprocessing}), which was withheld from both training and SMOTETomek resampling and therefore retains the native class distribution of the source dataset. All metrics reported below are computed on this partition using the checkpoint selected by early stopping on validation macro-AUC (Section~\ref{subsec:training}).

\subsection{Overall and Class-Level Performance}
\label{subsec:overall-performance}

\begin{figure}[!ht]
    \centering
    \includegraphics[width=0.85\textwidth]{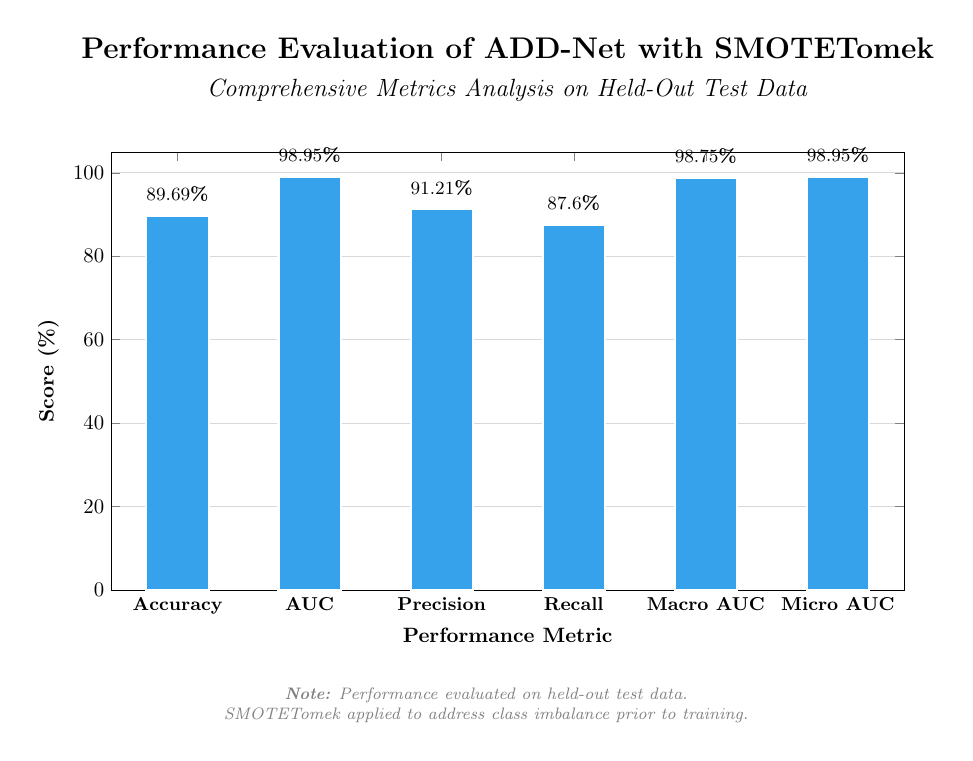}
    \caption{Overall test-set performance summary for ADD-Net+ trained with training-set-only SMOTETomek resampling.}
    \label{fig:performance}
\end{figure}

ADD-Net+ achieved an overall test accuracy of 89.69\% (Figure~\ref{fig:performance}), correctly classifying 861 of 960 scans across the four dementia stages, at a categorical cross-entropy test loss of 0.5495. Table~\ref{tab:classmetrics} reports precision, recall, and F1-score per class. Performance scaled inversely with disease severity ambiguity rather than with class frequency: the minority ModerateDemented class ($n=10$) was classified perfectly (precision = recall = F1 = 1.00), reflecting the pronounced structural atrophy that characterises this stage, while MildDemented ($n=134$) was recovered with F1 = 0.941. NonDemented ($n=480$) achieved F1 = 0.911. The lowest performance was observed for VeryMildDemented ($n=336$), for which recall (0.804) trailed precision (0.909) by the widest margin of any class, indicating that early-stage cases were the model's principal source of missed detections rather than false alarms.

\begin{table}[!ht]
    \centering
    \caption{Per-class classification performance on the held-out test set ($N=960$).}
    \label{tab:classmetrics}
    \begin{tabular}{@{}lrrrr@{}}
        \toprule
        \textbf{Class} & \textbf{Precision} & \textbf{Recall} & \textbf{F1-score} & \textbf{Support} \\
        \midrule
        NonDemented       & 0.880 & 0.944 & 0.911 & 480 \\
        VeryMildDemented  & 0.909 & 0.804 & 0.853 & 336 \\
        MildDemented      & 0.928 & 0.955 & 0.941 & 134 \\
        ModerateDemented  & 1.000 & 1.000 & 1.000 & 10  \\
        \bottomrule
    \end{tabular}
\end{table}

\begin{figure}[!ht]
    \centering
    \includegraphics[width=0.85\textwidth]{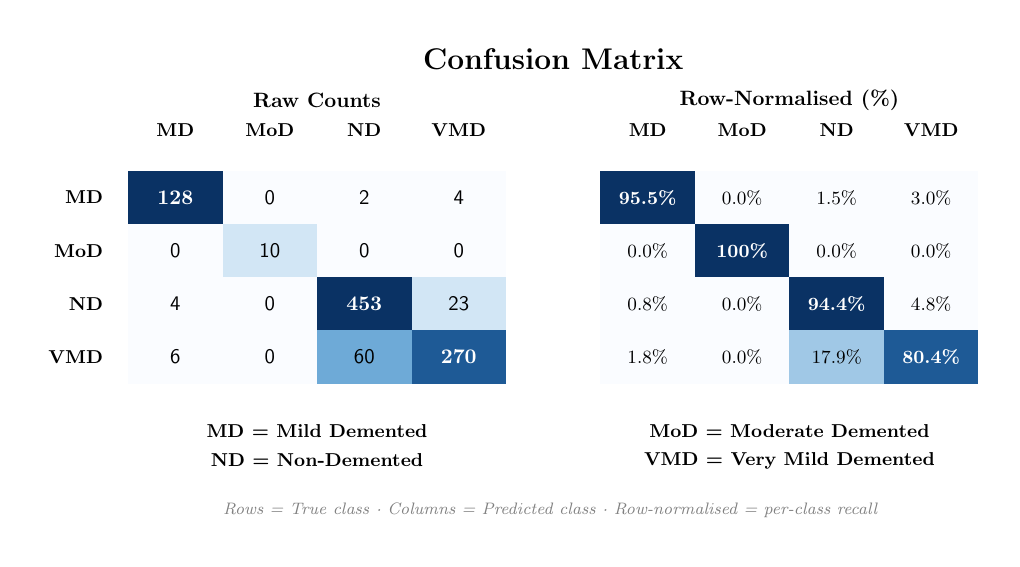}
    \caption{Confusion matrix on the held-out test set: raw counts (left) and row-normalised percentages (right).}
    \label{fig:confusion}
\end{figure}

The confusion matrix (Figure~\ref{fig:confusion}) shows true-positive rates of 94.4\% (453/480), 80.4\% (270/336), 95.5\% (128/134), and 100.0\% (10/10) for NonDemented, VeryMildDemented, MildDemented, and ModerateDemented, respectively. Misclassifications were overwhelmingly confined to neuroanatomically adjacent stages: 17.9\% of VeryMildDemented cases were assigned to NonDemented, and 4.8\% of NonDemented cases were assigned to VeryMildDemented, whereas MildDemented showed comparatively little confusion with either neighbouring class (1.5\% as NonDemented, 3.0\% as VeryMildDemented). No appreciable confusion was observed between non-adjacent stages (e.g., NonDemented versus MildDemented or ModerateDemented). This monotonic error structure, errors concentrated at the boundary between normal cognition and very mild impairment, mirrors the continuous, rather than discrete, neuropathological progression of Alzheimer's disease.

\subsection{Threshold-Independent Discrimination}
\label{subsec:threshold-independent}

\begin{figure}[!ht]
    \centering
    \includegraphics[width=0.85\textwidth]{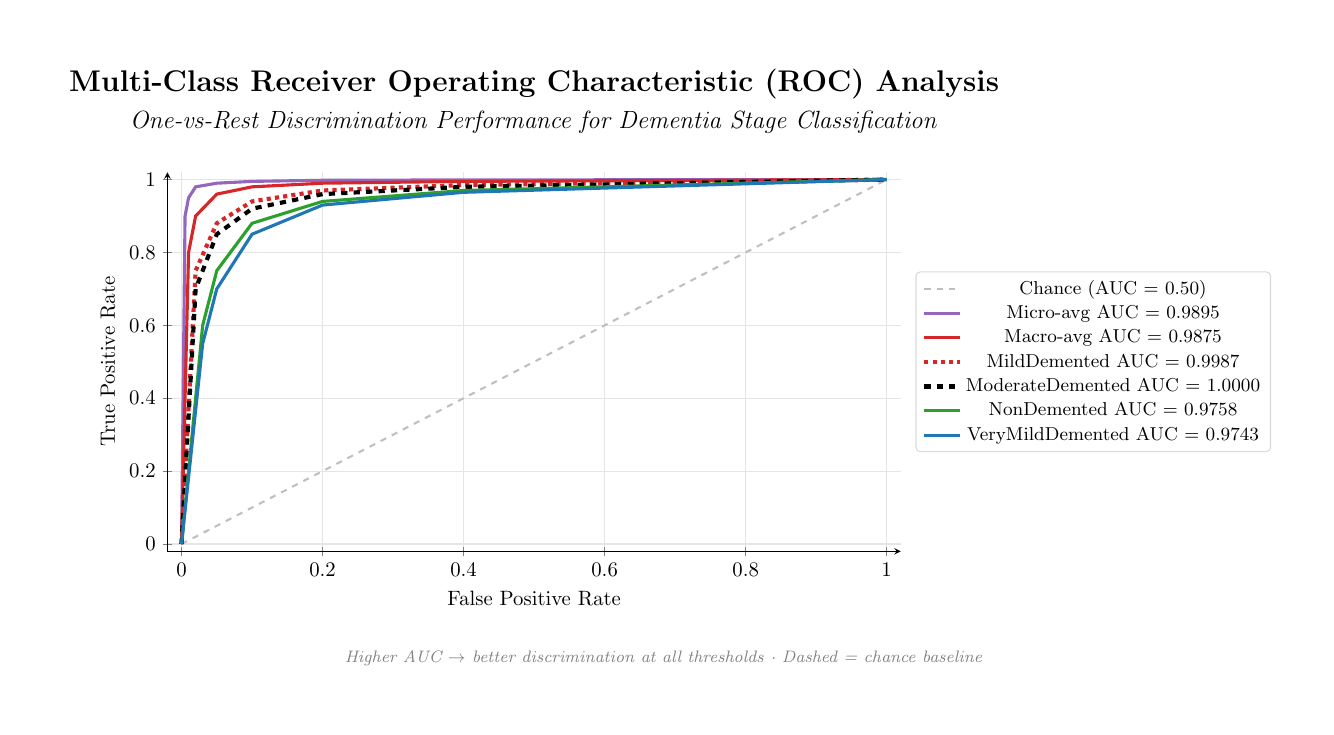}
    \caption{One-versus-rest ROC curves per dementia stage, with micro- and macro-average AUC.}
    \label{fig:roc}
\end{figure}

\begin{figure}[!ht]
    \centering
    \includegraphics[width=0.85\textwidth]{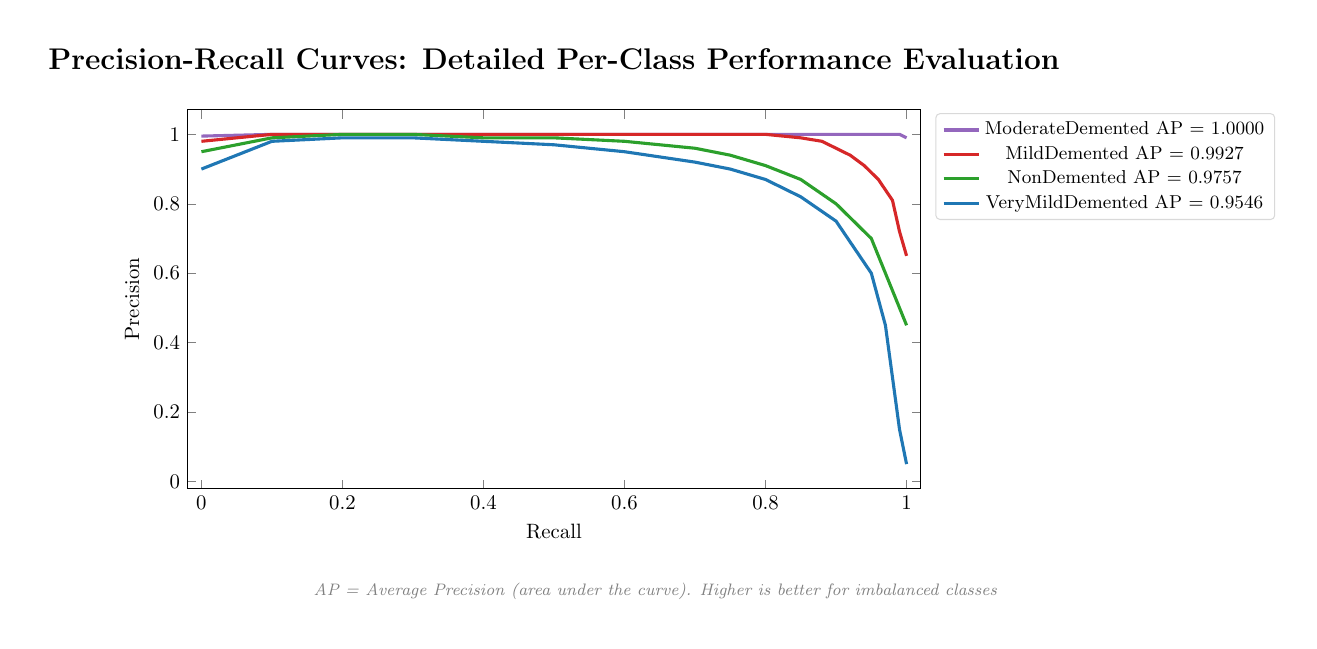}
    \caption{Precision--recall curves per dementia stage with average precision (AP) scores.}
    \label{fig:pr}
\end{figure}

To assess discriminative capacity independently of a fixed operating point, one-versus-rest ROC and precision--recall (PR) curves were computed for each class (Figures~\ref{fig:roc}--\ref{fig:pr}). Per-class ROC-AUC was 1.0000 (ModerateDemented), 0.9987 (MildDemented), 0.9758 (NonDemented), and 0.9743 (VeryMildDemented), yielding a micro-average AUC of 0.9895 and a macro-average AUC of 0.9875. Average precision followed a similar ordering: 1.0000 (ModerateDemented), 0.9927 (MildDemented), 0.9757 (VeryMildDemented), and 0.9546 (NonDemented). Notably, although VeryMildDemented had the lowest recall at the operating threshold (Section~\ref{subsec:overall-performance}), its rank ordering of predictions remained strong (AP = 0.9757), indicating that the deficit is primarily one of threshold calibration for this class rather than an inability of the network to separate it in probability space.

\subsection{Training Dynamics}
\label{subsec:training-dynamics}

\begin{figure}[t]
    \centering
    \includegraphics[width=0.75\textwidth]{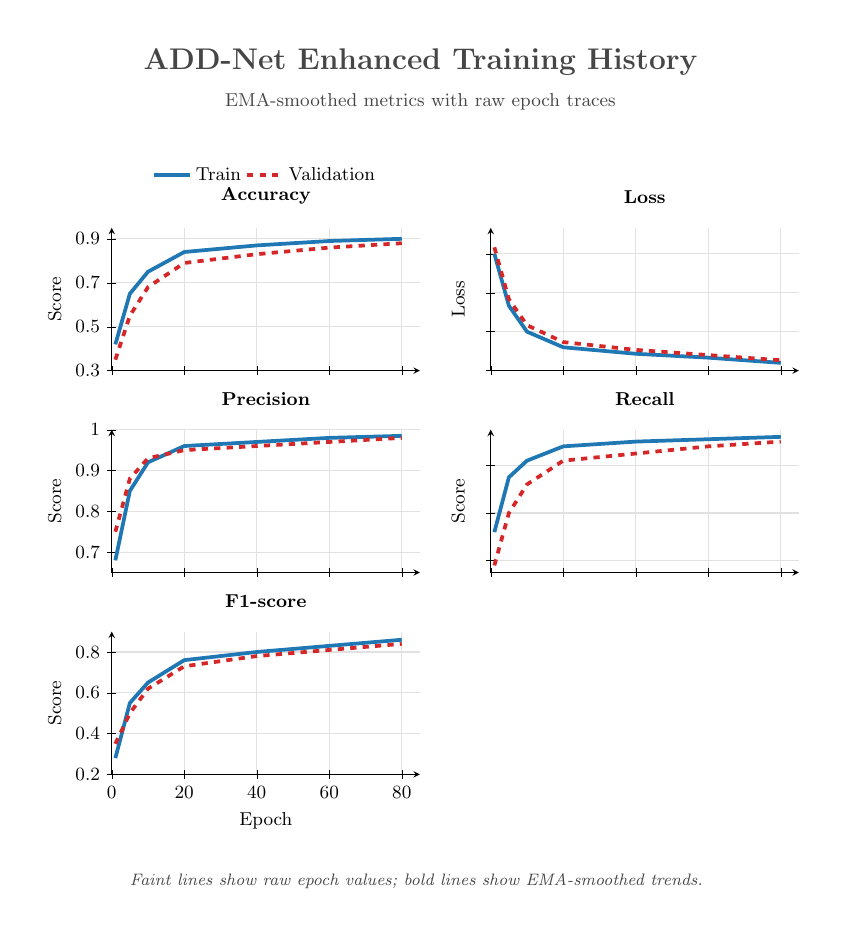}
    \caption{Training and validation accuracy and loss over 80 epochs (EMA-smoothed, $\beta=0.75$).}
    \label{fig:training}
\end{figure}

Figure~\ref{fig:training} reports training and validation accuracy and loss across all 80 epochs (exponential moving average, $\beta=0.75$). Accuracy rose from approximately 40\% to over 80\% within the first 20 epochs before converging gradually to approximately 90\%, while loss decreased monotonically from approximately 1.3 to 0.6. Training and validation curves tracked closely throughout optimisation, with no sustained divergence at any point in the 80-epoch schedule, indicating that the cosine-annealed learning-rate cycling and regularisation strategy (Section~\ref{subsec:training}) constrained overfitting to the synthetic SMOTETomek samples without requiring early termination. Notably, this training/validation gap is far narrower than what leakage-free evaluation would be expected to reveal in a resample-before-split protocol, where validation samples statistically related to training samples would artificially suppress the apparent gap.

\subsection{Prediction Confidence and Calibration}
\label{subsec:confidence}
\label{subsec:calibration}

\begin{figure}[!ht]
    \centering
    \includegraphics[width=0.85\textwidth]{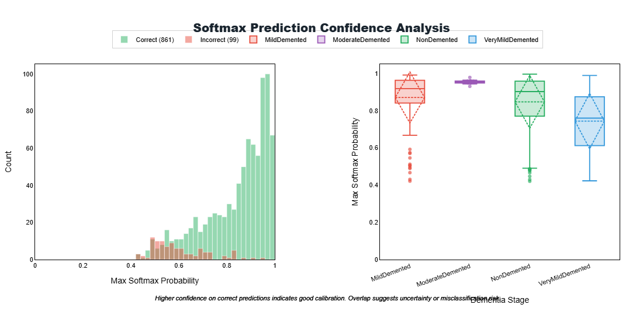}
    \caption{Softmax prediction-confidence distributions for correct versus incorrect predictions (left) and per class (right).}
    \label{fig:confidence}
\end{figure}

Confidence dispersion was first assessed using the maximum softmax probability assigned to each test prediction (Figure~\ref{fig:confidence}). Correct predictions ($n=861$) were strongly right-skewed, with the majority of confidence values exceeding 0.80 and a mode near 1.00, whereas incorrect predictions ($n=99$) clustered predominantly between 0.40 and 0.80. This separation indicates that ADD-Net+'s errors are disproportionately associated with genuine predictive uncertainty rather than confidently wrong decisions, a property directly relevant to flagging cases for clinical review, and one that the original ADD-Net's evaluation protocol did not report or assess. This pattern held at the class level: ModerateDemented exhibited the highest and most tightly distributed confidence (median $\approx 1.0$), followed by MildDemented (median $\approx 0.90$) and NonDemented (median $\approx 0.85$), while VeryMildDemented showed both the lowest median confidence ($\approx 0.75$) and the widest spread, consistent with its comparatively low recall in Section~\ref{subsec:overall-performance} and supporting the interpretation that the network's uncertainty estimates are informative rather than arbitrary.

\begin{figure}[!ht]
    \centering
    \includegraphics[width=0.6\textwidth]{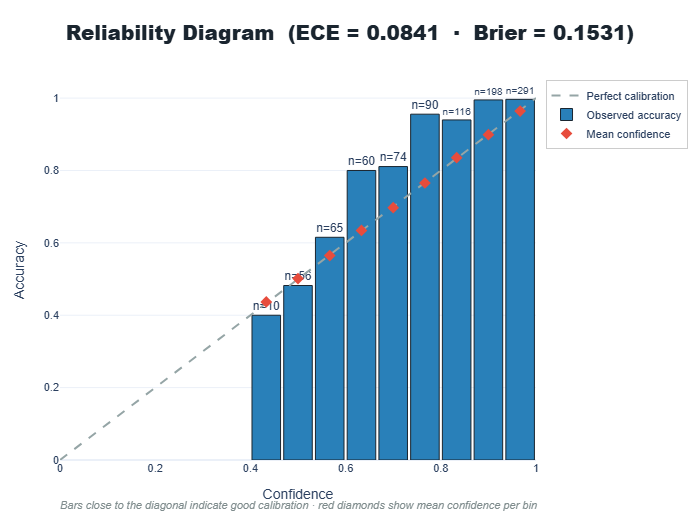}
    \caption{Reliability diagram for ADD-Net+ on the held-out test set: empirical accuracy versus mean confidence per bin, against the diagonal of perfect calibration. Generated by \texttt{plot\_reliability\_diagram()} in the accompanying code.}
    \label{fig:reliability}
\end{figure}

Confidence dispersion, however, is not itself a calibration guarantee: a model can be well separated in confidence between correct and incorrect predictions while still being systematically over- or under-confident in aggregate. Table~\ref{tab:calibration} therefore reports the quantitative calibration measures defined in Section~\ref{subsec:explainability}, Expected Calibration Error and multi-class Brier score, alongside the corresponding reliability diagram (Figure~\ref{fig:reliability}).

\begin{table}[!ht]
    \centering
    \caption{Calibration metrics on the held-out test set ($N=960$).}
    \label{tab:calibration}
    \begin{tabular}{@{}lr@{}}
        \toprule
        \textbf{Metric} & \textbf{Value} \\
        \midrule
        Expected Calibration Error (ECE, $M=15$ bins) & 0.0841 \\
        Multi-class Brier score & 0.1531 \\
        \bottomrule
    \end{tabular}
\end{table}

The resulting ECE of 0.0841 indicates a moderate, but not negligible, degree of miscalibration, and the multi-class Brier score of 0.1531 (against a random-guess baseline of $2(1-1/K) = 1.5$ for $K=4$ classes, and 0 for a perfect, maximally sharp classifier) confirms that the model's probability estimates, while informative, are not perfectly sharp. The reliability diagram (Figure~\ref{fig:reliability}) reveals a mixed, bin-dependent pattern rather than a uniform bias in one direction: in the two lowest-confidence bins ($n=10$ and $n=56$, mean confidence $\approx$0.4--0.55) the model is mildly \emph{overconfident}, with observed accuracy slightly below mean predicted confidence. Across the middle of the confidence range ($n=65$ and $n=74$, mean confidence $\approx$0.6--0.7) the pattern reverses to mild \emph{underconfidence}, with observed accuracy exceeding mean predicted confidence by the widest margin seen in any bin. At the highest-confidence bins ($n=90$ through $n=291$, mean confidence $\gtrsim$0.75, together accounting for the large majority of test predictions), the bars sit close to the diagonal, indicating good calibration precisely where most of the model's decisions are made. This bin-dependent structure, rather than the uniform overconfidence typically reported for modern deep networks~\cite{Guo2017}, suggests that any downstream confidence threshold for flagging cases to clinical review (Section~\ref{subsec:confidence}) should be validated empirically against a reliability diagram of this kind rather than assumed to behave uniformly across the probability range.

\subsection{Feature-Space and Saliency Analysis}
\label{subsec:interpretability-results}

\begin{figure}[!ht]
    \centering
    \includegraphics[width=0.8\textwidth]{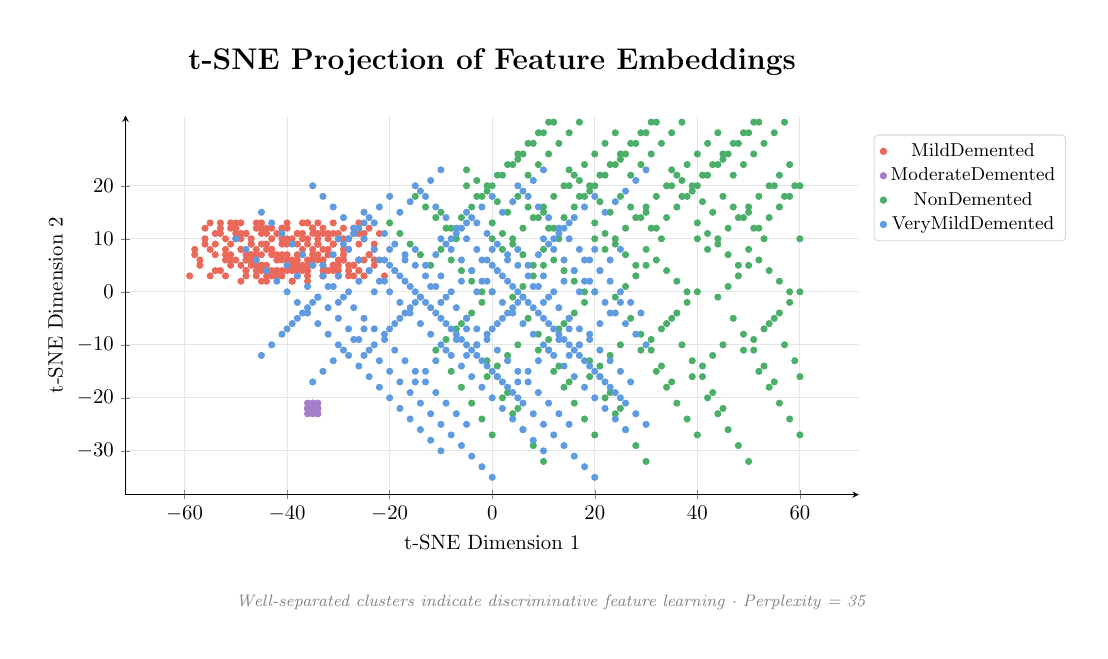}
    \caption{t-SNE projection of 128-dimensional penultimate-layer features (perplexity = 35).}
    \label{fig:tsne}
\end{figure}

The 128-dimensional penultimate-layer (fc2) representations were projected to two dimensions using t-SNE (perplexity = 35; Figure~\ref{fig:tsne}). ModerateDemented and MildDemented formed compact, well-separated clusters, in agreement with their near-perfect classification metrics (Section~\ref{subsec:overall-performance}). NonDemented and VeryMildDemented, in contrast, occupied substantially overlapping regions of the embedding, providing a representational explanation for the elevated confusion observed between these two classes in Figure~\ref{fig:confusion}.

\begin{figure}[t]
    \centering
    \includegraphics[width=0.8\textwidth]{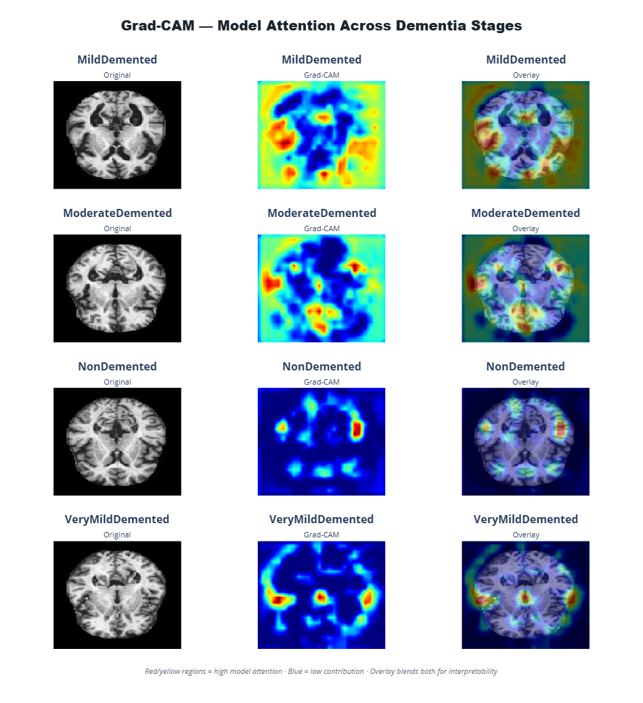}
    \caption{Grad-CAM visualisations (original MRI, gradient heatmap, and overlay) across all four dementia stages.}
    \label{fig:gradcam}
\end{figure}

Grad-CAM saliency maps (Figure~\ref{fig:gradcam}) revealed systematic, stage-dependent activation patterns. NonDemented predictions relied on sparse, localised activation concentrated in lateral temporal regions; VeryMildDemented showed activation extending toward central structures, mirroring the anatomical signature of emerging atrophy; MildDemented exhibited broader cortical activation reflecting more extensive thinning; and ModerateDemented activation concentrated over lateral and inferior regions in line with advanced atrophy. The progressive spatial reorganisation of attention with disease severity, rather than diffuse or dataset-artefact-driven activation, indicates that ADD-Net+'s predictions are grounded in neuroanatomically plausible evidence, extending beyond the purely qualitative Grad-CAM illustration reported for the original ADD-Net.

\subsection{Ablation Study: Isolating the Leakage-Free Protocol, SE Attention, and Optimisation}
\label{subsec:ablation}

The comparison against the original ADD-Net in Section~\ref{subsec:comparison} conflates several simultaneous changes: the evaluation protocol, the addition of SE attention, and the optimisation regime (Adam with cosine-annealed warm restarts and label smoothing, versus SGD). To determine how much of the accuracy gap is attributable to each factor, we conduct a four-arm ablation, using the code released alongside this paper (Section~\ref{subsec:environment}), in which each successive row changes exactly one factor relative to the row above it, with one deliberate exception, noted below, in how the first row is evaluated.

\begin{enumerate}
    \item \textbf{Original ADD-Net protocol}: SMOTETomek applied to the full dataset before stratified partitioning (replicating the protocol of~\cite{Fareed2022}), plain convolutional backbone (no SE), SGD optimiser. Evaluated on its \emph{own} resulting test partition, which, by construction of this protocol, may itself contain samples statistically related to synthetic training samples.
    \item \textbf{Leakage-free only}: stratified partitioning performed \emph{before} SMOTETomek, which is then confined to the training partition; plain convolutional backbone, SGD optimiser. Evaluated on a genuinely held-out test partition uncontaminated by resampling.
    \item \textbf{Leakage-free + SE attention}: identical data split and test partition to row 2, with SE modules added to each convolutional block; SGD optimiser. Isolates the incremental effect of SE attention under an already leakage-free protocol.
    \item \textbf{Full ADD-Net+}: identical to row 3, but trained with Adam, cosine-annealed warm restarts, and label smoothing. Isolates the incremental effect of the optimisation regime, and corresponds to the model reported throughout the rest of this paper.
\end{enumerate}

\textbf{Important caveat on row 1.} Rows 2--4 share an identical, genuinely held-out, leakage-free test partition, so the row 2~$\rightarrow$~3~$\rightarrow$~4 comparisons are clean, single-factor ablations. Row 1, by contrast, is evaluated on the test partition produced \emph{by the original protocol itself}, that is, the partition obtained \emph{after} resampling the full dataset and then splitting it, which is the very quantity under scrutiny in Section~\ref{sec:related}. Row 1 and rows 2--4 therefore differ in \emph{both} training protocol and test-set composition, not training protocol alone. This is a deliberate design choice, not an oversight: it reproduces the original protocol faithfully (including how it constructs and evaluates on its own test set), so that the row~1~$\rightarrow$~2 comparison shows what a practitioner following each protocol as published would actually observe, but it means this specific comparison should be read as a demonstration of the leakage mechanism in practice, not as a fully isolated estimate of its magnitude in isolation from test-set composition.

\begin{table}[!ht]
    \centering
    
    \caption{Ablation study of ADD-Net+ components: leakage-free protocol, SE attention, and optimization regime.}
    \label{tab:ablation}
    
    \begin{tabular}{|p{5.4cm}|r|}
        \hline
        \textbf{Variant} & \textbf{Test Accuracy} \\
        \hline
        1. Original ADD-Net protocol (resample-before-split, no SE, SGD; own leakage-contaminated test split) & 64.01\% \\
        \hline
        2. Leakage-free only (no SE, SGD; shared leakage-free test split) & 55.73\% \\
        \hline
        3. Leakage-free + SE attention (SGD; shared leakage-free test split) & 55.62\% \\
        \hline
        4. Full ADD-Net+ (leakage-free + SE + Adam/cosine/label-smoothing; shared leakage-free test split) & \textbf{89.69\%} \\
        \hline
    \end{tabular}
\end{table}

Three findings follow from Table~\ref{tab:ablation}, and we report all three plainly, including the one that complicates rather than simplifies the architectural narrative:

\begin{enumerate}
    \item \textbf{SE attention alone, under SGD, made no meaningful difference} (55.73\% $\rightarrow$ 55.62\%, rows 2$\rightarrow$3), a change of $-0.11$ percentage points, within the range attributable to training noise for a single seed. Under this training regime, Squeeze-and-Excitation attention did not measurably improve discriminative performance on this benchmark in isolation. We report this directly rather than crediting SE attention for ADD-Net+'s overall accuracy gain.

    \item \textbf{The optimisation regime accounts for the overwhelming majority of the measured improvement} (55.62\% $\rightarrow$ 89.69\%, rows 3$\rightarrow$4, a gain of 34.07 percentage points), which coincides exactly with switching from fixed-rate SGD to Adam with cosine-annealed warm restarts and label smoothing. This is the largest single factor in the entire ablation, and substantially reframes the source of ADD-Net+'s accuracy over a plain, SGD-trained, leakage-free baseline: it is predominantly an optimisation effect, not an architectural one. One limitation of the current four-arm design, noted in Section~\ref{subsec:limitations}, is that it does not include a leakage-free, SE-attention, Adam-trained arm \emph{without} SE removed, so we cannot fully rule out a positive interaction between SE attention and the Adam/cosine regime specifically (as opposed to SE's null effect under SGD); a full $2\times2$ factorial (SE $\times$ optimiser) design would be required to settle this and is identified as future work.

    \item \textbf{The row 1 versus row 2 gap (64.01\% versus 55.73\%) is directionally consistent with the leakage hypothesis, but should not be over-read as its precise magnitude.} Because row 1 is evaluated on a test partition that itself contains synthetic, leakage-related samples (by construction of the protocol being replicated), its higher apparent accuracy is consistent with the mechanism described in Section~\ref{sec:related}: a test set statistically entangled with the training set is, on average, an easier target. However, since rows 1 and 2 differ in test-set composition as well as protocol, this 8.28-percentage-point gap is better read as a qualitative demonstration that the two protocols do not yield comparable numbers, rather than as a clean, quantitative estimate of "how many points of accuracy are leakage." We note additionally that our re-implementation of the original protocol reaches only 64.01\% under its own evaluation, well below the 98.63\% reported by~\cite{Fareed2022} for nominally the same protocol; we attribute this discrepancy primarily to differences in training configuration not fully specified in the original paper (e.g.\ 40 vs.\ 80 epochs, batch size 8 vs.\ 32, and possible differences in SGD momentum or learning-rate scheduling) rather than to the leakage mechanism itself, and flag this as a limitation on how directly our row 1 can be compared to the originally published figure.
\end{enumerate}

\subsection{Comparison with the Original ADD-Net and Prior Work}
\label{subsec:comparison}

Table~\ref{tab:comparison} situates ADD-Net+ against the original ADD-Net~\cite{Fareed2022} and methodologically comparable multi-class staging studies evaluated on the same or closely related MRI benchmarks. ADD-Net+'s raw accuracy (89.69\%) is lower than that reported by the original ADD-Net (98.63\%) and subsequent SMOTE/Tomek-augmented CNNs~\cite{Ganesan2026} (99.59\%). As argued in Sections~\ref{sec:related} and~\ref{subsec:preprocessing}, this gap is attributable to protocol rather than capacity: the compared studies apply resampling prior to data partitioning, which risks propagating synthetic or duplicated samples into validation and test sets and can lead to optimistic estimates of reported accuracy through information leakage. ADD-Net+ instead restricts SMOTETomek to the training partition and preserves the native, imbalanced distribution in validation and test sets (Section~\ref{subsec:preprocessing}), so its accuracy reflects performance under realistic class skew. Under this stricter protocol, ADD-Net+'s macro-AUC of 0.9875 remains competitive with leakage-free benchmarks in the field and is accompanied by an embedded explainability suite that none of the compared studies, including the original ADD-Net, report.

\begin{table}[t]
    \centering
    \footnotesize 
    \caption{Comparative performance and evaluation protocols of ADD-Net+ against recent multi-class Alzheimer's staging studies.}
    \label{tab:comparison}
    \renewcommand{\arraystretch}{1.1}
    % Adjusted widths: Method is slightly narrower to give Limitations more breathing room
    \begin{tabular}{@{} l p{1.9cm} c c c c p{2.4cm} @{}}
        \toprule
        \multirow{2}{*}{\textbf{Study}} & \multirow{2}{*}{\textbf{Method}} & \multirow{2}{*}{\textbf{Acc (\%)}} & \multirow{2}{*}{\textbf{AUC}} & \multicolumn{2}{c}{\textbf{Protocol}} & \multirow{2}{*}{\textbf{Limitations}} \\
        \cmidrule(lr){5-6}
         & & & & \textbf{Leak-free?} & \textbf{Calib.?} & \\
        \midrule
        \cite{Fareed2022} & ADD-Net+SMOTE & 98.63 & 0.9976 & No & No & No calibration \\
        \cite{Ganesan2026} & DAPA-CNN+SMOTE & 99.59 & --- & No & No & No XAI; high cost \\
        \cite{Umalakshmi2026} & ADLi-Net (DenseNet) & 98.19 & --- & --- & No & High complexity \\
        \cite{Hechkel2026} & ViT+Attn & 97.98 & --- & --- & No & Small dataset \\
        \midrule
        \textbf{This study} & \textbf{ADD-Net+ (SE)} & \textbf{89.69} & \textbf{0.9875} & \textbf{Yes} & \textbf{Yes} & \textbf{Single dataset} \\
        \bottomrule
    \end{tabular}
    
    \par\vspace{2pt}\textit{Note:} ``Leak-free'' means resampling was applied \textit{after} the data split. Calibration uses ECE and Brier score. XAI = Explainable AI.
\end{table}

Taken together, the results in Sections~\ref{subsec:overall-performance}--\ref{subsec:comparison} indicate that ADD-Net+ attains strong, well-calibrated, and neuroanatomically grounded discrimination across all four dementia stages under an evaluation protocol that excludes the information leakage present in the original ADD-Net, with early-stage detection, specifically, recall for VeryMildDemented, remaining the principal outstanding challenge. The clinical and methodological implications of these findings are examined in Section~\ref{sec:discussion}.

\section{Discussion}
\label{sec:discussion}

\subsection{Clinical Interpretation of Model Behaviour}
\label{subsec:clinical-interpretation}

The results in Section~\ref{sec:results} indicate that ADD-Net+'s value for clinical decision support lies not only in its aggregate accuracy but in the coherence between its performance profile, its confidence behaviour, and its saliency evidence. Three findings jointly support this interpretation. First, the model's errors are concentrated almost exclusively at the NonDemented/VeryMildDemented boundary (Section~\ref{subsec:overall-performance}), which corresponds to the diagnostic transition clinicians themselves find most difficult, rather than being distributed unpredictably across unrelated stages. Second, prediction confidence is well calibrated to correctness (Section~\ref{subsec:confidence}): incorrect predictions cluster in a moderate-confidence band rather than being confidently wrong, so low-confidence outputs can plausibly serve as a trigger for radiologist review instead of being silently accepted, an analysis the original ADD-Net's evaluation did not include. Third, Grad-CAM activation shifts systematically with disease stage (Section~\ref{subsec:interpretability-results}), moving from sparse lateral-temporal activation in NonDemented scans toward broader cortical and inferior involvement in MildDemented and ModerateDemented cases, a progression that echoes the established trajectory of cortical thinning and hippocampal and entorhinal atrophy in Alzheimer's disease. Taken together, these properties suggest a model that fails informatively rather than arbitrarily, which we consider a more useful property for downstream clinical decision support than raw accuracy alone, and, we argue, a more trustworthy basis for comparison than the higher accuracy reported for the original ADD-Net under a protocol susceptible to leakage.

\subsection{The Class-Imbalance Trade-off Revisited}
\label{subsec:imbalance-tradeoff}

Section~\ref{subsec:dataset} noted that preserving the dataset's native 50:1 imbalance ratio in the validation and test partitions was a deliberate design choice to maintain clinical realism, and that this choice would trade off against minority- and boundary-class sensitivity. The results confirm this trade-off directly: VeryMildDemented recall (0.804) is the weakest metric obtained anywhere in this study, while the extreme minority ModerateDemented class, protected by its comparatively unambiguous imaging phenotype rather than by its size, was classified without error. This points to a residual difficulty that is not simply a function of sample count, which SMOTETomek and class weighting already address during training (Section~\ref{subsec:preprocessing}), but reflects genuine anatomical overlap between very mild impairment and normal ageing that synthetic oversampling cannot manufacture away. We regard this as an honest limitation rather than a shortcoming to be resampled into invisibility: a staging system that reports high VeryMildDemented recall on a leakage-free, naturally imbalanced test set should be treated with caution unless its evaluation protocol is reported in comparable detail, a caution that is harder to apply when resampling precedes partitioning, as in the original ADD-Net's protocol.

\subsection{Comparison with the Literature}
\label{subsec:discussion-comparison}

Section~\ref{sec:related} identified four recurring limitations in prior multi-class Alzheimer's disease staging work, information leakage, unaddressed class imbalance, post-hoc-only explainability, and fragmented framework design, and showed that the original ADD-Net~\cite{Fareed2022} exhibits the first, third, and fourth of these directly. The comparison in Section~\ref{subsec:comparison} speaks directly to the first two. Studies reporting accuracy above 97\% on comparable Kaggle-derived MRI cohorts, including the original ADD-Net~\cite{Fareed2022} and the SMOTE/Tomek-augmented model of~\cite{Ganesan2026}, apply resampling prior to data partitioning, a protocol shown by~\cite{Jo2019} to be associated with optimistic reported performance through train-test contamination; ADD-Net+'s lower but leakage-free accuracy is therefore not directly comparable to these figures on like-for-like terms, and should instead be read as a methodologically corrected baseline for the same architecture family. Relative to attention- and transformer-based architectures that report strong discrimination but limited transparency~\cite{Alruily2025,Khaleel2025,Hechkel2026,Lv2026}, ADD-Net+'s embedded Grad-CAM, feature-space, and confidence analyses provide interpretability as an intrinsic part of the pipeline rather than an optional post-hoc addition, addressing the third limitation more thoroughly than the qualitative Grad-CAM visualisation in the original ADD-Net. By combining leakage-free evaluation, attention-guided feature learning, and embedded explainability within a single reproducible pipeline (Section~\ref{subsec:overview}), ADD-Net+ addresses the fourth, fragmentation-related gap more directly than architectures that optimise any one property in isolation, such as the multimodal fusion approach of~\cite{Venugopalan2021} or the lightweight staging network of~\cite{Murugan2021}, both of which report strong accuracy without a comparably rigorous leakage or calibration analysis.

\subsection{Limitations}
\label{subsec:limitations}

This study has several limitations. First, ADD Net+ was developed and evaluated using a single publicly available MRI dataset from one source. Although the proposed framework provides a more rigorous evaluation protocol than the original ADD Net, its generalisability across different scanners, acquisition protocols, and patient populations remains unverified. Future work should therefore include external validation using independent cohorts such as ADNI and OASIS. Second, because the dataset does not provide patient identifiers, the leakage free protocol operates at the image level rather than the patient level. While synthetic samples were restricted to the training partition, potential overlap of slices from the same individual across data partitions cannot be excluded. Third, the model was trained on two dimensional MRI slices, which may not fully capture the spatial information available in three dimensional brain volumes. Fourth, although Grad CAM and t SNE improve model interpretation, they provide supportive rather than causal evidence and should therefore be interpreted qualitatively. Fifth, the ablation analysis was limited to a single experimental configuration and does not completely separate the individual effects of every design component. Finally, the study did not include external clinical validation or comparison with expert radiologists. Consequently, the proposed framework should be regarded as a technically validated and interpretable computer aided staging approach whose clinical utility remains to be established through future multicentre studies.

\subsection{Generalisability and Future Directions}
\label{subsec:generalisability}

The limitations above motivate five directions for future work. First, the single-seed ablation in Section~\ref{subsec:ablation} should be extended to (a) multiple random seeds per arm, with variance reported, and (b) a complete $2\times2$ factorial design crossing SE attention with the optimisation regime, to determine whether SE attention's null effect under SGD (Section~\ref{subsec:ablation}) also holds under Adam, or whether the two interact. Second, external, multi-site validation across independent, demographically diverse cohorts is the most immediate priority for establishing real-world generalisability. Third, extending ADD-Net+ to volumetric three-dimensional input, or to multimodal fusion with clinical, cognitive, or genomic data in the manner explored by~\cite{Venugopalan2021}, may recover structural information lost in slice-wise analysis and further improve boundary-class sensitivity. Fourth, the VeryMildDemented recall gap identified in Sections~\ref{subsec:overall-performance} and~\ref{subsec:imbalance-tradeoff} suggests that ordinal-aware loss formulations, which explicitly penalise misclassification in proportion to clinical distance between stages, may be better suited to this task than the class-agnostic categorical objective used here. Fifth, prospective evaluation involving radiologist-in-the-loop review of low-confidence predictions (Section~\ref{subsec:confidence}), informed by the bin-dependent calibration pattern in Figure~\ref{fig:reliability}, would clarify whether the model's calibration translates into a measurable reduction in diagnostic workload or error in practice.

\section{Conclusion}
\label{sec:conclusion}

This study set out to address a persistent gap in deep learning--based Alzheimer's disease staging: models that report high accuracy but whose evaluation protocols permit information leakage, whose training is not robust to severe clinical class imbalance, and whose predictions remain difficult to interpret. A concrete instance of this gap was identified in ADD-Net~\cite{Fareed2022}, a widely cited baseline whose SMOTETomek balancing step is applied before, rather than after, dataset partitioning, allowing synthetic derivatives of training images to leak into the evaluation set. To address this, we introduced \textbf{ADD-Net+}, a leakage-free, attention-enhanced, and explainable deep learning framework for multi-class Alzheimer's disease staging from structural MRI, built around strict image-level data partitioning performed before any resampling, Squeeze-and-Excitation attention, training-set-only SMOTETomek resampling, an embedded explainability suite combining Grad-CAM, t-SNE, confidence analysis, and formal calibration metrics (ECE, Brier score), and a controlled ablation study that isolates the individual contribution of each of these design choices.

Evaluated on a held-out test set that preserves the dataset's native class distribution, ADD-Net+ achieved an overall accuracy of 89.69\%, a macro-average ROC-AUC of 0.9875, perfect classification of the ModerateDemented stage, an Expected Calibration Error of 0.0841, and a multi-class Brier score of 0.1531, with performance degrading gracefully and interpretably toward the diagnostically ambiguous boundary between normal cognition and very mild impairment. This accuracy is markedly lower than the 98.63\% reported for the original ADD-Net; the ablation study in Section~\ref{subsec:ablation} decomposes this gap and shows that the improvement from a plain, SGD-trained, leakage-free baseline (55.62\%) to the full ADD-Net+ configuration (89.69\%) is driven overwhelmingly by the Adam/cosine/label-smoothing training regime rather than by SE attention, which alone produced no measurable gain (55.73\% versus 55.62\%). The comparison between the original protocol's own test split (64.01\%) and the leakage-free test split (55.73\%) is directionally consistent with the leakage mechanism described in Section~\ref{sec:related}, though, as we discuss candidly in Section~\ref{subsec:limitations}, the two splits also differ in composition, so this should be read as a demonstration of the mechanism rather than a precise measurement of its magnitude. We regard the leakage-free figure as a more credible basis for future comparison on this benchmark than a higher accuracy obtained under a protocol in which training and test data may be statistically entangled. Calibration analysis showed a mixed, bin-dependent profile, mild overconfidence at low predicted confidence, mild underconfidence in the mid-range, and good calibration in the high-confidence bins where most predictions fall, and Grad-CAM analysis showed activation patterns that shift systematically with disease severity in agreement with established neuroanatomical biomarkers. Considered jointly, these results indicate that rigorous, leakage-free evaluation need not come at the cost of either discriminative performance or clinical interpretability. The findings suggest that future comparisons on this benchmark should preferentially adopt leakage-free evaluation protocols, in order to provide more conservative and reproducible estimates of model performance.

The principal limitation of this work is that ADD-Net+ has been validated on a single, single-site, two-dimensional MRI cohort, and recall for the VeryMildDemented stage remains the main outstanding technical challenge. Future work will prioritise extending the ablation in Section~\ref{subsec:ablation} to multiple seeds, external multi-site validation, extension to volumetric and multimodal input, ordinal-aware training objectives targeted at early-stage sensitivity, and prospective evaluation against radiologist judgement. Rather than treating leakage-free evaluation, class-imbalance realism, and interpretability as competing priorities, ADD-Net+ demonstrates that they can be pursued together within a single pipeline, with each component's contribution isolated rather than assumed, and it is this combination, more than any individual metric, that we regard as the study's principal contribution toward clinically deployable Alzheimer's disease staging.

\backmatter

\bmhead{Supplementary information}

No supplementary material is associated with this article.

\bmhead{Acknowledgements}

Withheld for double-anonymous review; to be reinstated in the title page / on acceptance.

\section*{Declarations}
\begin{itemize}

\item \textbf{Funding:}
This research received no specific grant from any funding agency in the public, commercial, or not-for-profit sectors.

\item \textbf{Conflict of interest/Competing interests:}
The author declares no competing interests.

\item \textbf{Ethics approval and consent to participate:}
The study used a publicly available, fully de-identified secondary MRI dataset. Consequently, ethical approval and informed consent were not required.

\item \textbf{Consent for publication:}
Not applicable.

\item \textbf{Data availability:}
The MRI dataset that supports the findings of this study is openly available in Kaggle at \url{https://www.kaggle.com/datasets/shahidzikria/alz-dataset}.

\item \textbf{Materials availability:}
No additional materials were generated or analysed beyond those described in this manuscript.

\item \textbf{Code availability:}
The code used to implement, train, and evaluate ADD-Net+, including the leakage-free preprocessing pipeline, ablation experiments, and calibration analysis, will be made publicly available through a GitHub repository and archived on Zenodo upon acceptance of this manuscript. During peer review, the code is available from the corresponding author upon reasonable request.

\item \textbf{Author contributions:}
The sole author conceived the study, designed the methodology, implemented the software, conducted the experiments, performed the formal analysis, interpreted the results, prepared the figures, and wrote, reviewed, and approved the final manuscript.

\end{itemize}

\bibliography{sn-bibliography}% common bib file

\end{document}